\documentclass[letterpaper,english,reprint,superscriptaddress,longbibliography]{revtex4-1}
\usepackage{lmodern}

\usepackage[T1]{fontenc}
\usepackage[utf8]{inputenc}
\usepackage{babel}
\usepackage{units}
\usepackage{amsmath}
\usepackage{amssymb}
\usepackage{graphicx}
\usepackage[unicode=true,pdfusetitle,
 bookmarks=false,
 breaklinks=false,pdfborder={0 0 0},pdfborderstyle={},backref=false,colorlinks=false]
 {hyperref}

\makeatletter

\pdfpageheight\paperheight
\pdfpagewidth\paperwidth

\makeatother

\begin{document}

\title{Single photons and unconventional photon blockade in quantum dot
cavity-QED}

\author{H.J. Snijders}

\affiliation{Huygens-Kamerlingh Onnes Laboratory, Leiden University, P.O. Box
9504, 2300 RA Leiden, The Netherlands}

\author{J. A. Frey}

\affiliation{Department of Physics, University of California, Santa Barbara, California
93106, USA}

\author{J. Norman}

\affiliation{Department of Electrical \& Computer Engineering, University of California,
Santa Barbara, California 93106, USA}

\author{H. Flayac}

\affiliation{Institute of Physics iPHYS, École Polytechnique Fédérale de Lausanne
EPFL, CH-1015 Lausanne, Switzerland}

\author{V. Savona}

\affiliation{Institute of Physics iPHYS, École Polytechnique Fédérale de Lausanne
EPFL, CH-1015 Lausanne, Switzerland}

\author{A. C. Gossard}

\affiliation{Department of Electrical \& Computer Engineering, University of California,
Santa Barbara, California 93106, USA}

\author{J. E. Bowers}

\affiliation{Department of Electrical \& Computer Engineering, University of California,
Santa Barbara, California 93106, USA}

\author{M. P. van Exter}

\affiliation{Huygens-Kamerlingh Onnes Laboratory, Leiden University, P.O. Box
9504, 2300 RA Leiden, The Netherlands}

\author{D. Bouwmeester}

\affiliation{Huygens-Kamerlingh Onnes Laboratory, Leiden University, P.O. Box
9504, 2300 RA Leiden, The Netherlands}

\affiliation{Department of Physics, University of California, Santa Barbara, California
93106, USA}

\author{W. Löffler}

\affiliation{Huygens-Kamerlingh Onnes Laboratory, Leiden University, P.O. Box
9504, 2300 RA Leiden, The Netherlands}
\begin{abstract}
We observe the unconventional photon blockade effect in quantum dot
cavity QED, which, in contrast to conventional photon blockade, operates
in the weak coupling regime. A single quantum dot transition is simultaneously
coupled to two orthogonally polarized optical cavity modes, and by
careful tuning of the input and output state of polarization, the
unconventional photon blockade effect is observed. We find a minimum
second-order correlation $g^{(2)}(0)\approx0.37$ which corresponds
to $g^{(2)}(0)\approx0.005$ when corrected for detector jitter, and
observe the expected polarization dependency and photon bunching and
anti-bunching very close-by in parameter space, which indicates the
abrupt change from phase to amplitude squeezing.
\end{abstract}
\maketitle
\textbf{}

A two-level system strongly coupled to a cavity results in polaritonic
dressed states with a photon-number dependent energy. This dressing
gives rise to the photon blockade effect \cite{Imamoglu1997,Milburn1989}
resulting in photon-number dependent transmission and reflection,
enabling the transformation of incident coherent light into specific
photon number states such as single photons. Single-photon sources
are a crucial ingredient for various photonic quantum technologies
ranging from quantum key distribution to optical quantum computing.
Such sources are characterized by a vanishing second-order auto-correlation
$g^{(2)}(0)\approx0$ \cite{senellart2017}. 

In the strong coupling regime, where the coupling between the two-level
system and the cavity is larger than the cavity decay rate $(g>\kappa)$
\cite{Loudon-book}, photon blockade has been demonstrated in atomic
systems \cite{Birnbaum2005}, quantum dots in photonic crystal cavities
\cite{Faraon2008}, and circuit QED \cite{Lang2011a,Hoffman2011}.
At the onset of the weak coupling regime $(g\approx\kappa)$, it has
been shown that by detuning the dipole transition frequency with respect
to the cavity resonance, photon blockade can still be observed \cite{Muller2014}.
However, moving further into the weak coupling regime $(g<\kappa)$
which is much easier to achieve, conventional photon blockade is no
longer possible because the energy gap between the polariton states
disappears. Nevertheless, also in the weak coupling regime, the two-level
system enables photon number sensitivity, which has recently enabled
high-quality single photon sources using polarization postselection
\cite{Somaschi2015a,Ding,Snijders2017} or optimized cavity in-coupling
\cite{DeSantis2017}\footnote{We focus here only on resonantly excited systems.}.

\begin{figure}
\includegraphics[scale=0.4]{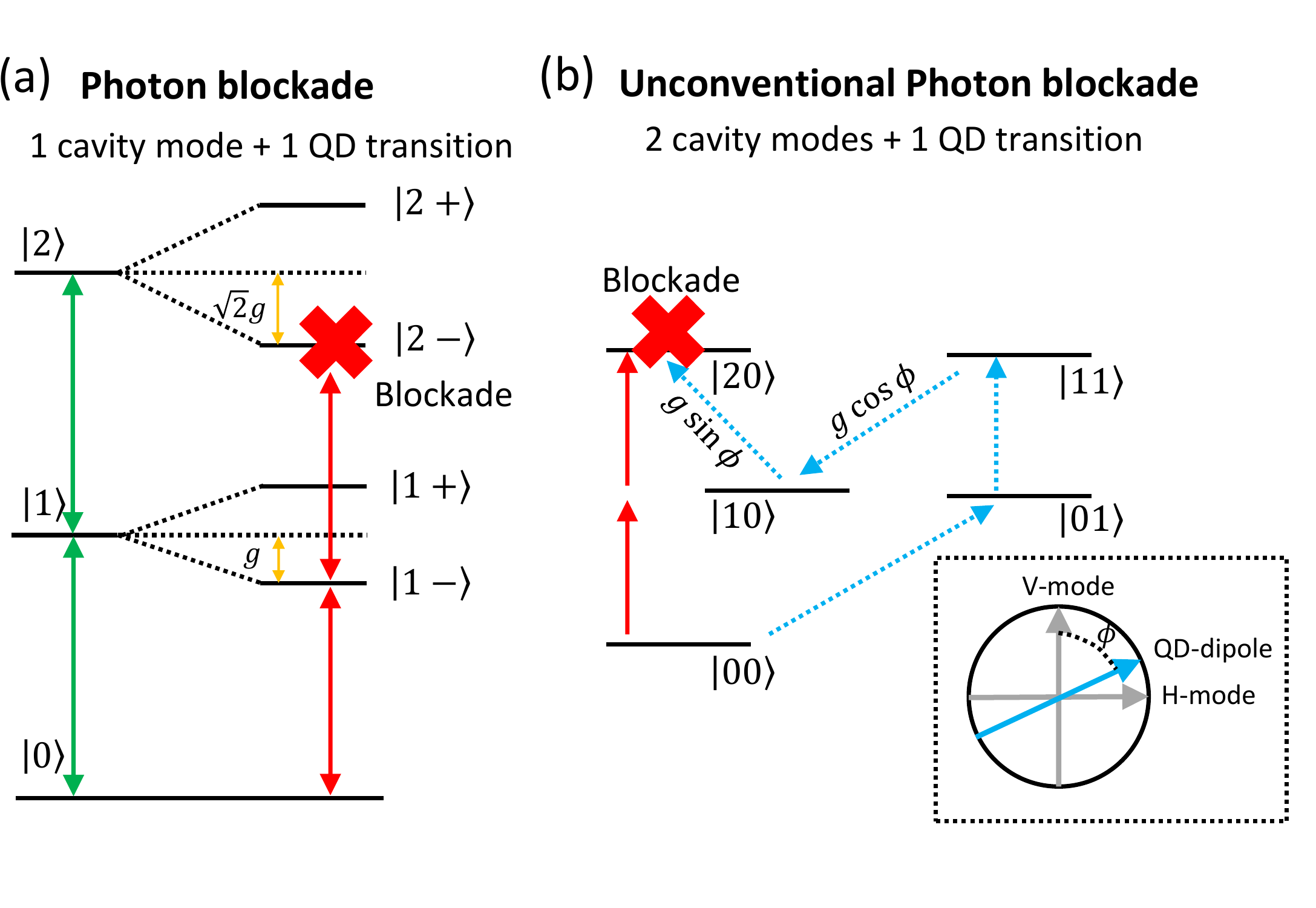}

\caption{Removal of the 2-photon component in conventional photon blockade
by the anharmonicity of the Jaynes Cummings ladder (a). In the unconventional
photon blockade (b, adapted from \cite{Bamba2011}), two excitation
pathways (red and blue arrows) destructively interfere. The state
$\left|ij\right\rangle $ corresponds to $(i,j)$ photons in the ($H,V$)
polarized micro cavity modes. The quantum dot is coupled (coupling
constant $g$) to both cavity modes due to an orientational mismatch
of its dipole (angle $\phi$, see inset). \label{fig:Sketch}}
\end{figure}

In 2010, Liew and Savona introduced the concept of the unconventional
photon blockade effect \cite{Liew2010} which operates even for weak
non-linearities. It is a quantum interference effect between different
excitation pathways which requires at least two or more degrees of
freedom \cite{Flayac2017}. It was first investigated for Kerr non-linearities
\cite{Liew2010,Flayac2013}, then for $\chi^{(2)}$ non-linearities
\cite{Gerace2014b} and the Jaynes Cummings \cite{Bamba2011} system
which we focus on here. Both the conventional and unconventional photon
blockade effect result in transmitted light with vanishing photon
auto-correlation $g^{(2)}(0)<10^{-2}$ \cite{Verger2006,Flayac2013},
however, the underlying physical mechanisms are completely different,
see Fig. \ref{fig:Sketch}. In the strong coupling regime, the nonlinearity
of the dressed spectrum prevents reaching the two photon state for
a particular laser frequency {[}red arrows in Fig. \ref{fig:Sketch}(a){]},
which is impossible in the weak coupling regime (green arrows). In
unconventional photon blockade {[}Fig. \ref{fig:Sketch}(b){]}, reduction
of the two-photon state is achieved by destructive interference between
different excitation pathways \cite{Bamba2011,Wang2017}. We investigate
here a single semiconductor quantum dot in an optical micro cavity
where a single linearly polarized quantum dot dipole transition is
coupled to the two linearly polarized cavity modes due to an orientational
mismatch of the quantum dot dipole with respect to the cavity axes
{[}angle $\phi$, see inset Fig. \ref{fig:Sketch}(b){]}. Since the
unconventional photon blockade operates better in the low mean photon
number regime, Fig. \ref{fig:Sketch}(b) shows only the $N=0$...$2$
photon Fock states. Further, we show only one particular excitation
pathway (blue), many more involving internal cavity coupling exist
but do not qualitatively change the interpretation: By tuning the
input and output superposition between the two cavity modes (here,
via polarization), the interference of different excitation pathways
with and without involvement of the photon-number sensitive quantum
dot transition can be tuned such that the two-photon component vanishes.

In this paper, we show experimental evidence of unconventional photon
blockade (UPB) in quantum dot cavity-QED. The sample consist of a
layer of self-assembled InAs/GaAs quantum dots embedded in a micropillar
cavity (maximum Purcell factor $F_{p}=11.2$) grown by molecular beam
epitaxy \cite{Strauf2007a}. The quantum dot layer is embedded in
a P–I–N junction, separated by a 27 nm thick tunnel barrier from the
electron reservoir to enable tuning of the quantum dot resonance frequency
by the quantum-confined Stark effect. Due to the quantum dot fine-structure
structure splitting, we need to consider only one quantum dot transition,
which interacts with both the H and V cavity modes.

We model our system using a Jaynes-Cummings Hamiltonian in the rotating
wave approximation with $g\ll\kappa.$ The Hamiltonian for two cavity
modes and one quantum dot transition driven by a continuous wave laser
is written as
\begin{align*}
H & =\left(\omega_{L}-\omega_{c}^{V}\right)\hat{a}_{V}^{\dagger}\hat{a}_{V}+\left(\omega_{L}-\omega_{c}^{H}\right)\hat{a}_{H}^{\dagger}\hat{a}_{H}\\
 & +\left(\omega_{L}-\omega_{QD}\right)\hat{\sigma}^{\dagger}\hat{\sigma}+g\left(\hat{\sigma}\hat{b}^{\dagger}+\hat{\sigma}^{\dagger}\hat{b}\right)\\
 & +\eta_{H}\left(\hat{a}_{H}+\hat{a}_{H}^{\dagger}\right)+\eta_{V}\left(\hat{a}_{V}+\hat{a}_{V}^{\dagger}\right).
\end{align*}

$\omega_{c}^{H}$ and $\omega_{c}^{V}$ are the resonance frequencies
of the linearly polarized cavity modes, $\hat{a}_{H}^{\dagger}$ and
$\hat{a}_{V}^{\dagger}$ the photon creation operators, $\omega_{QD}$
is the quantum dot resonance frequency, and $\hat{\sigma}^{\dagger}$
the exciton creation operator. $\hat{b}=\hat{a}_{V}\cos\,\phi+\hat{a}_{H}\sin\,\phi$
is the cavity photon annihilation operator along the quantum dot dipole
orientation, and $\phi$ is the relative angle. In our case the angle
is $\phi=94^{\circ}$, which means that the H-cavity mode couples
better to the exciton transition. $\eta_{H}$ and $\eta_{V}$ are
the amplitudes of the incident coherent light coupling to the H and
V cavity modes. For numerical simulations, we add relaxation of the
cavity modes and dephasing of the quantum dot transition and solve
the corresponding quantum master equation \cite{Johansson2012a,Johansson2013,Snijders2016,Snijders2017},
add the output polarizer and calculate the mean photon number and
second order correlation function. All theoretically obtained $g^{(2)}(\tau)$
data is convoluted with the detector response (530 ps) to match the
experimental setup.

\begin{figure}
\includegraphics[scale=0.55]{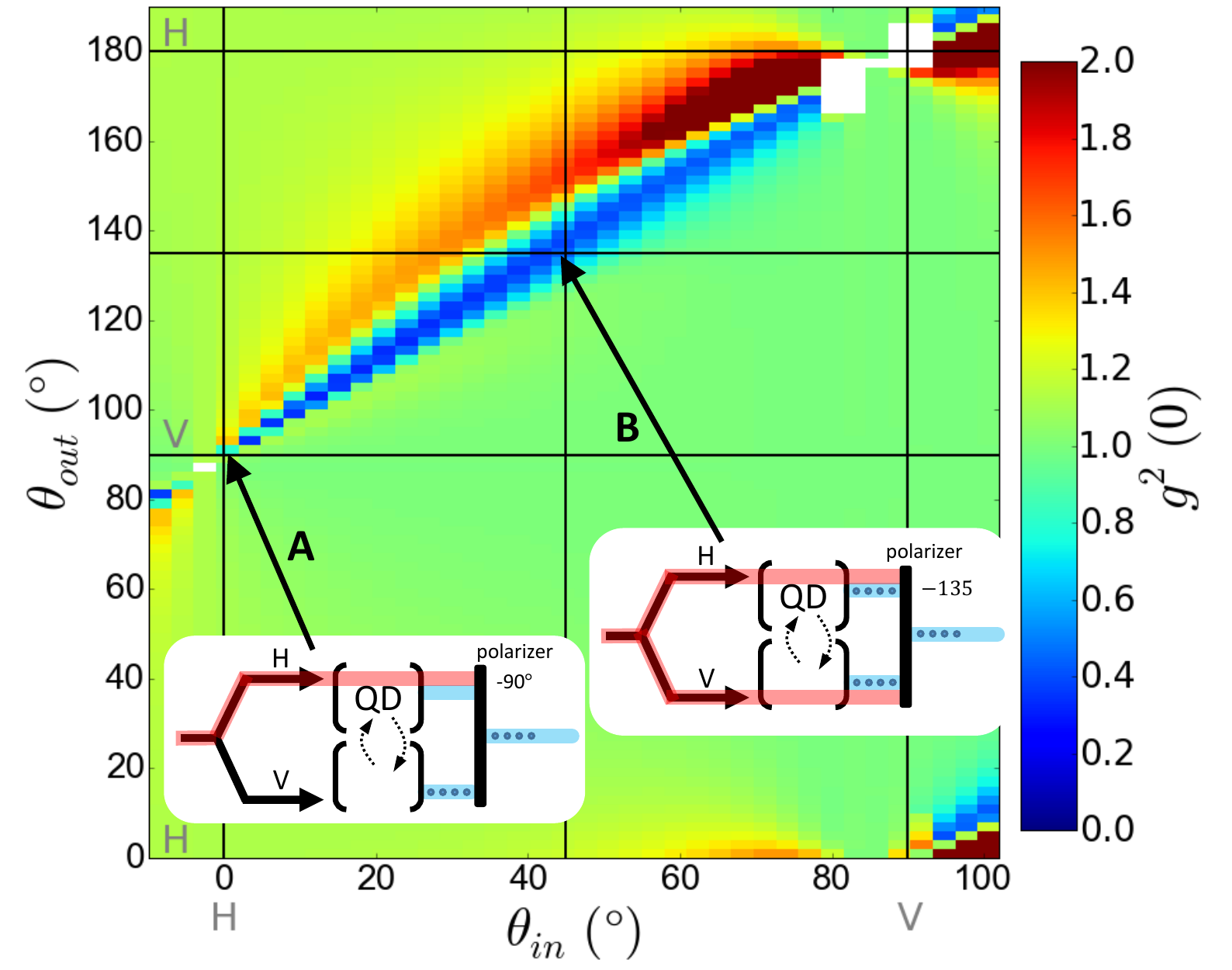}

\caption{False color plot of the theoretically calculated $g^{(2)}(0)$ convoluted
with the detector response as a function of the incident and detected
linear polarization orientation. Arrow A indicates the condition where
most single photon sources operate: the system is excited in the H-cavity
mode and the single photons are detected in the V-cavity mode. Arrow
B\textbf{ }shows the case where single photons are created using the
unconventional photon blockade. White pixels indicate that the simulation
has failed due to extremely low photon numbers. \label{fig:False-color-plots} }
\end{figure}

Fig. \ref{fig:False-color-plots} shows how the second order correlation
$g^{(2)}(\tau=0)$ of the transmitted photons depends on the linear
input and linear output polarization angle. In all current quantum
dot based single photon sources \cite{Somaschi2015a,Ding,Snijders2017},
only one cavity mode is excited with the laser, and by using a crossed
polarizer, single photons are obtained. This condition is indicated
with arrow A in Fig. \ref{fig:False-color-plots}. We see that, by
using an input polarization where both polarization modes are excited
(indicated by arrow B), a much larger range is found where single
photons can be produced. This is where the unconventional photon blockade
can be observed. 

\begin{figure}
\includegraphics[width=8.5cm]{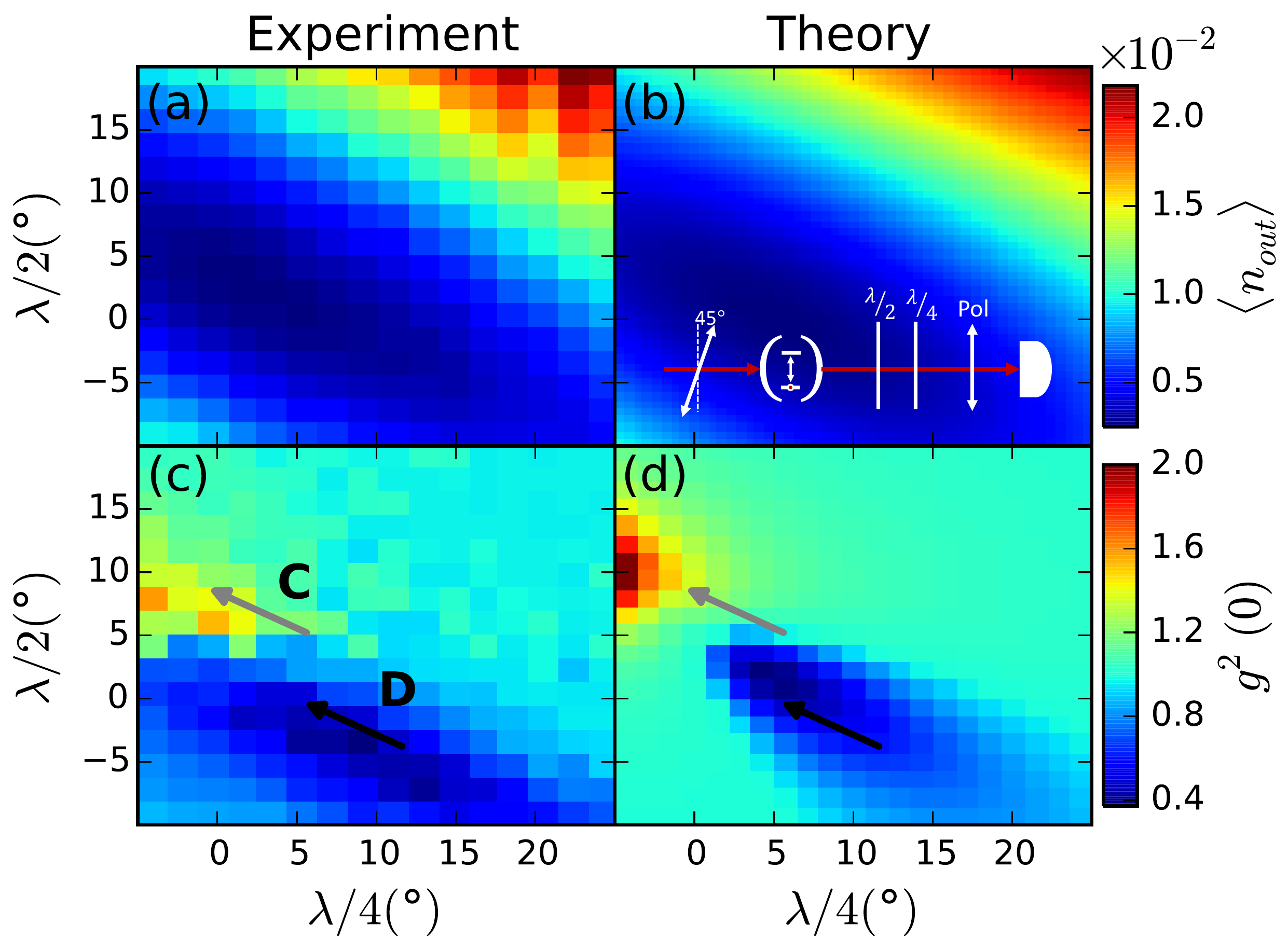}

\includegraphics[width=8.5cm]{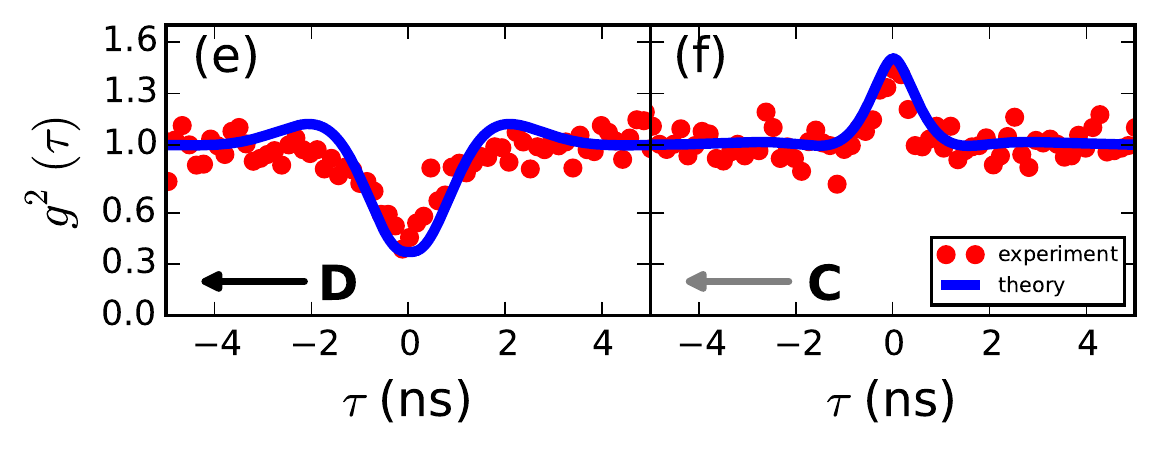}

\caption{False color plots of $\left\langle n_{out}\right\rangle $ and $g^{2}(0)$
as a function of the orientation of the $\nicefrac{\lambda}{2}$ and
$\nicefrac{\lambda}{4}$ wave plate in the transmission path. (a)
$\left\langle n_{out}\right\rangle $ is the mean photon number in
a given polarization basis at the output. At $0^{\circ}$ the linear
polarized incoming light is parallel to the fast axis of both wave
plates. (b) corresponding theory to (a) with as inset a sketch of
the experimental setup. (c) and (d) experimental and theoretical $g^{(2)}(0)$.
(e) and (f) show $g^{(2)}(\tau)$ for the (anti) bunching region indicated
by arrows C (D) in Fig \ref{g2(tau) plots}(c) and (d). The red dots
are measured data and the blue line is the theoretically obtained
$g^{(2)}(\tau)$ convoluted with the detector response. \label{g2(tau) plots}}
\end{figure}

Now, we investigate more closely region B of Fig. \ref{fig:False-color-plots},
where both cavity modes are excited ($\theta_{in}=45^{\circ}$). Furthermore,
we add the experimentally unavoidable polarization splitting of the
$H$ and $V$ cavity modes which is $10$ GHz for the device under
investigation. Furthermore, we vary the detected output polarization
in the most general way, by introducing $\lambda/2$ and $\lambda/4$
wave plates before the final polarizer in the transmission path. The
experimental setup is sketched in the inset of Fig. \ref{g2(tau) plots}(b).
Fig. \ref{g2(tau) plots}(b)\textbf{ }shows how this polarization
projection affects the mean photon number $\left\langle n_{out}\right\rangle $,
for $\left\langle n_{in}\right\rangle =\left(\frac{\eta_{H}+\eta_{V}}{\kappa}\right)^{2}=0.06$
in the simulation and in the experiment {[}Fig. \ref{g2(tau) plots}(a){]}.
This region is highly dependent on the cavity splitting and the quantum
dot dipole angle, careful determination of the parameters allows us
to obtain good agreement to experimental data {[}Fig. \ref{g2(tau) plots}(a){]}.
In this low mean photon number region, the second-order correlation
$g^{(2)}(0)$ shows a non-trivial behavior as a function of the output
polarization state, shown in Fig. \ref{g2(tau) plots}(c, experiment)
and (d, theory): First, we observe the expected unconventional photon
blockade anti-bunching (blue region). The experimentally measured
minimum $g^{(2)}(0)$ is $0.37$, which is limited by the detector
response function. The theoretical data which takes the detector response
into account agrees very well to the experimental data and predicts
a bare $g^{(2)}(0)\approx0.005$. Second, we find that, close-by in
parameter space, there is a region where bunched photons are produced.
This enhancement of the two-photon probability happens via constructive
interference leading to phase squeezing. Theoretical and experimental
data show good agreement, we attribute the somewhat more extended
antibunching region to long-time drifts during the course of the experiment
(10 hours). 

In Fig. \ref{g2(tau) plots}(e) and \ref{g2(tau) plots}(f) we show
the two-time correlation function $g^{(2)}(\tau)$ for the two cases
indicated by the arrows. The observed width and height of the anti-bunching
and bunching peak predicted by the theory is in excellent agreement
with the observed experimental data. For two coupled Kerr resonators
in the UPB regime, one observes oscillations in $g^{(2)}(\tau)$ when
collecting the output of only one of the cavities \cite{Liew2010}.
During finalizing this paper, a manuscript describing a first observation
of this effect has appeared \cite{vaneph2018}. In our case, these
oscillations are absent because the system works mostly as a unidirectional
dissipative coupler \cite{Flayac2016}, and the photon field behind
the output polarizer contains contributions from both cavities modes,
which suppresses the oscillations in $g^{(2)}(\tau)$.

An alternative way to understand the unconventional photon blockade
is in terms of Gaussian squeezed states \cite{Lemonde2014}: For any
coherent state $\left|\alpha\right\rangle $, there exists an optimal
squeeze parameter $\xi$ that minimizes the two-photon correlation
$g^{(2)}(0)$, which can be made vanishing for a weak driving fields.
We find that, even with a small amount of squeezing, it is possible
to significantly reduce the 2-photon distribution and minimize $g^{(2)}(0)$
for low mean photon numbers. A Gaussian squeezed state is produced
from vacuum like $D(\alpha)S(\xi)\left|0\right\rangle =\left|\alpha,\xi\right\rangle $.
Here $S$ is the squeeze operator with $\xi=r\exp^{i\theta}$ $(0\leq r<\infty$,
$0\leq\theta\leq2\pi)$. $D$ is the displacement operator, and the
complex displacement amplitude $\alpha=\bar{\alpha}\exp^{i\vartheta}$
($0\leq\bar{\alpha}<\infty$, $0\leq\vartheta\leq2\pi).$ For $\theta=\vartheta=0$,
we can calculate the two photon probability in the small-$\alpha$
(low mean photon number) limit as
\begin{equation}
\left|\left\langle 2\right|D(\alpha)S(\xi)\left|0\right\rangle \right|^{2}\approx(\bar{a}{}^{2}-r)^{2}/2,\label{prob-2-photon state}
\end{equation}

using a Taylor expansion. We see that, in order to obtain a vanishing
two-photon probability, the squeeze parameter $r$ needs to be equal
to $\bar{a}{}^{2}$ which is the mean photon number. By defining the
amount of quadrature squeezing as $\left\langle \left(\Delta X_{1}\right)^{2}\right\rangle =\frac{1}{4}e^{-2r}$
and considering a $\left\langle n_{out}\right\rangle \approx0.004$
(Fig. \ref{g2(tau) plots}(a)), this condition leads to $10\log_{10}(e^{-0.008})=-3\times10^{-2}$
dB squeezing. Interestingly, this result means that, for a weak coherent
state, only a very small amount of squeezing is needed to make $g^{(2)}(0)$
drop to zero. 

\begin{figure}
\includegraphics[width=8.5cm]{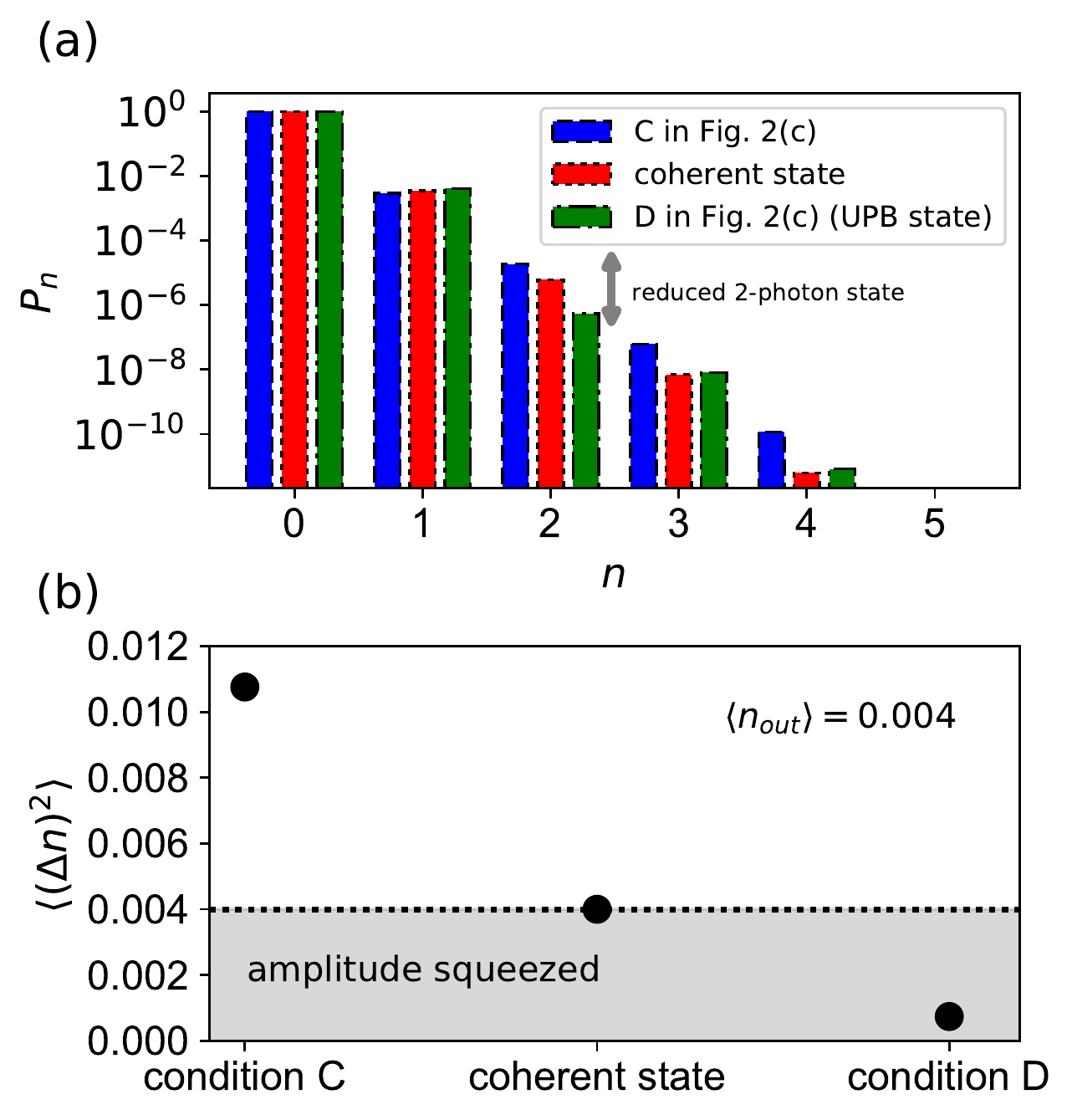}

\includegraphics[width=8.5cm]{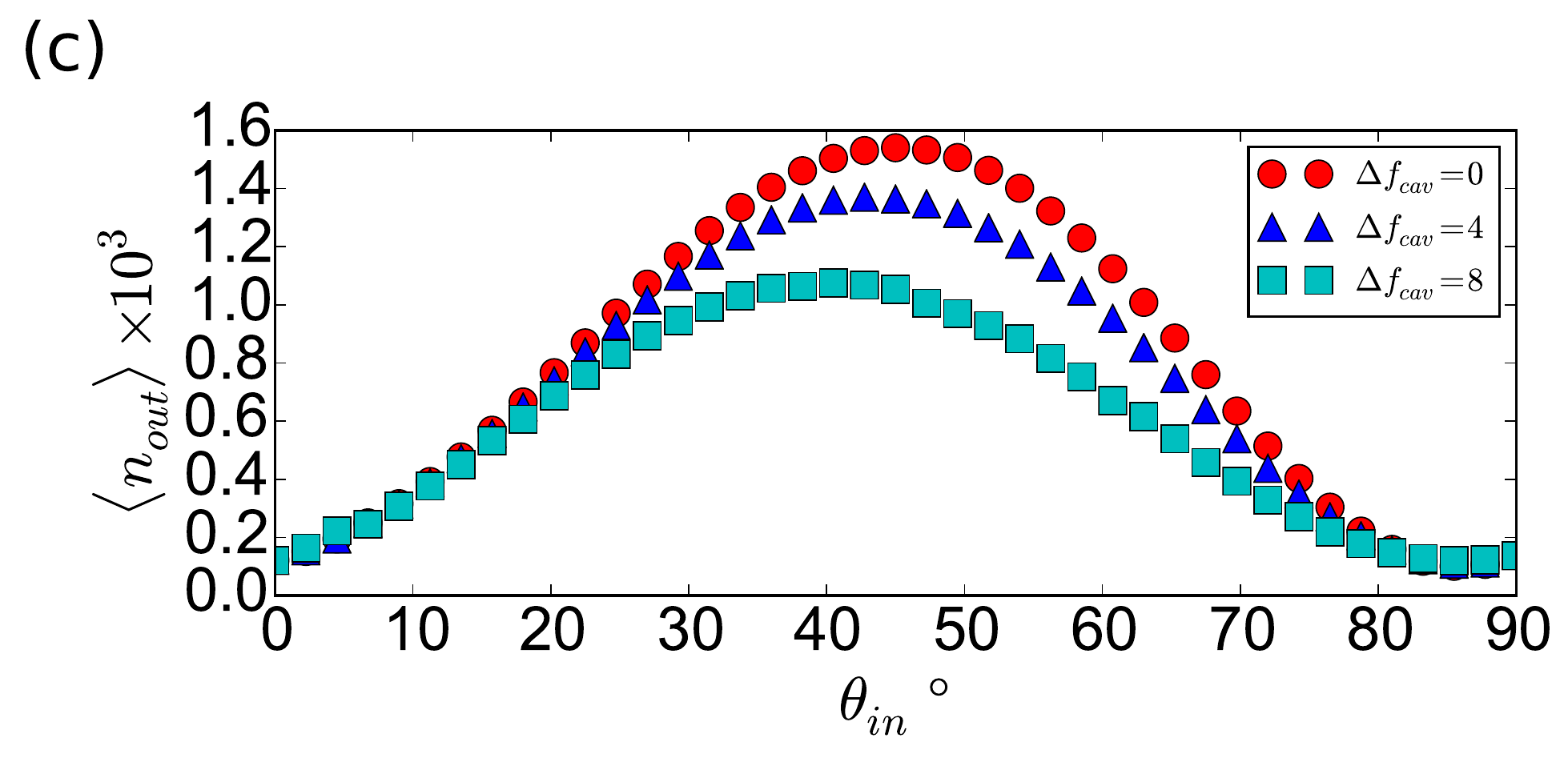}

\caption{(a) Calculated photon number distribution of a coherent state and
for the condition indicated by the arrow C and D in Fig. \ref{fig:False-color-plots}(c).
(b)\textbf{ }The calculated photon number variance for the states
presented in (a) showing amplitude squeezing in the region where we
observe the unconventional photon blockade. (c) Mean photon number
$\left\langle n_{out}\right\rangle $ as a function of input polarization.
We see that a large improvement of the single photon brightness can
be obtained by exploiting the UPB effect. The simulation is performed
for three cavity splittings $(\Delta f_{cav})$ showing that the enhancement
is largest in a polarization degenerate cavity. \label{UPB-Pn} }
\end{figure}

In Fig. \ref{UPB-Pn} we show further analysis of the theoretical
calculations for the experimental state produced by the unconventional
photon blockade as indicated by arrow D in Fig. \ref{fig:False-color-plots}(c)
and (d). In agreement with equation (\ref{prob-2-photon state}) we
observe that the 2-photon state in the photon number distribution
shown in Fig. \ref{UPB-Pn}(a) is suppressed. Consequently, from the
photon number variance given in Fig. \ref{UPB-Pn}(b), we observe
that the state is amplitude squeezed. Further, by moving from the
region of arrow C to D in Fig. \ref{g2(tau) plots}(c), the observed
state switches from a phase squeezed to an amplitude squeezed state,
which is a clear signature of the unconventional photon blockade effect
\cite{Flayac2017}.

Finally, we discuss whether the UPB effect can be used to enhance
the performance of single photon sources, and in particular their
brightness. Traditionally, the quantum dot is excited by one linearly
polarized cavity mode and photons are collected via the orthogonal
mode. In our experiment, the quantum dot excitation probability is
$1-\cos(4^{\circ})\approx0.0024$, and, once excited, it has $1-0.0024$
chance to emit into the collection cavity mode, which leads to a low
total efficiency. In the unconventional photon blockade regime, arrow
B in Fig \ref{fig:False-color-plots}, this efficiency is higher.
To further explore this, we show in Fig. \ref{UPB-Pn}(c) the mean
photon number $\left\langle n_{out}\right\rangle $ as a function
of the input polarization with constant input laser power (the polarization
output state is chosen such that $g^{(2)}(0)\approx0$). We see that,
by rotating the input polarization from $0^{\circ}$ to $45^{\circ}$,
the output mean photon number can be increased by approximately a
factor 10. The simulation is done for various cavity splittings $\Delta f_{cav}$
which shows that increasing the cavity splitting reduces this enhancement.

In conclusion, we have experimentally observed the unconventional
photon blockade effect using a single quantum dot resonance coupled
to two orthogonally polarized cavity modes. We find the expected drop
in $g^{(2)}(0)$, but additionally and very close in parameter space,
we also find that the transmitted light statistics can be tuned from
anti-bunched to bunched, all in good agreement to theoretical models
and simulations. In contrast to conventional photon blockade, no energy
splitting of the polariton resonances is required, allowing to obtain
$g^{(2)}(0)\approx0$ even with weak non-linearities. Finally, under
certain conditions, we find that the unconventional photon blockade
effect can increase the brightness of the single photon sources. 
\begin{acknowledgments}
We thank D. Kok and M.F. Stolpe for fruitful discussions. We acknowledge
funding from the Netherlands Organisation for Scientific Research
(NWO) (Grant No. 08QIP6-2), from NWO and the Ministry for Education,
Culture and Science (OCW) as part of the Frontiers of Nanoscience
program, and from the National Science Foundation (NSF) (0901886,
0960331).
\end{acknowledgments}

\bibliographystyle{naturemagwV1allauthors}
\bibliography{Mybibliography-squeez2}

\end{document}